\documentclass[pra,twocolumn,showpacs,superscriptaddress]{revtex4-2}
\usepackage{times}
\usepackage{latexsym}
\usepackage{graphicx}
\usepackage{amsmath,amssymb}
\usepackage{verbatim,times,bbm}
\usepackage{soul}
\usepackage[colorlinks,hyperindex]{hyperref}
\hypersetup{colorlinks,citecolor=blue,linkcolor=blue,urlcolor=blue}
\usepackage{color}

\UseRawInputEncoding

\usepackage{epsf,epsfig}
\usepackage{ifpdf}
\usepackage[T1]{fontenc}
\usepackage{graphicx}
\usepackage{epsfig}
\usepackage{epstopdf}
\usepackage{arcs}
\usepackage{subfigure}
\usepackage{tensor}
\usepackage{braket}
\usepackage{float}
\usepackage[dvipsnames]{xcolor}
\usepackage{amsmath}
\usepackage{morefloats}
\newcommand{\be}{\begin{equation}}
	\newcommand{\ee}{\end{equation}}

\usepackage{tikz}
\usetikzlibrary{calc}   
\usetikzlibrary{topaths}
\usepackage{xcolor}
\usepackage{hyperref}
\hypersetup{
	colorlinks =WildStrawberry,
	linkcolor =red,
	pdftitle={Leading corrections to Kerr geometry},
	citecolor=NavyBlue,
	filecolor=WildStrawberry,
	urlcolor=WildStrawberry,}
\usepackage[toc]{appendix}
\usepackage{color,soul}
\usepackage{datetime}

\newcommand{\bea}{\begin{eqnarray}}
	\newcommand{\eea}{\end{eqnarray}}
\newcommand{\ba}{\begin{eqnarray}}
	\newcommand{\ea}{\end{eqnarray}}

\newcommand{\beq}{\begin{equation}}
	\newcommand{\eeq}{\end{equation}}
\newcommand{\beqa}{\begin{eqnarray}}
	\newcommand{\eeqa}{\end{eqnarray}}
\newcommand{\beqar}{\begin{eqnarray*}}
	\newcommand{\eeqar}{\end{eqnarray*}}

\usepackage{hyperref}
\hypersetup{
	colorlinks =WildStrawberry,
	linkcolor =red,
	pdftitle={Leading corrections to Kerr geometry},
	citecolor=NavyBlue,
	filecolor=WildStrawberry,
	urlcolor=WildStrawberry, }
\begin{document}
\title{Geometric phase of a two-level atom near a dielectric nanosphere out of thermal equilibrium}

\author{Ehsan Amooghorban\footnote{ehsan.amooghorban@sku.ac.ir}}
\affiliation{Department of Physics,  Faculty of Science,  Shahrekord University P. O. Box 115,  Shahrekord ,  Iran}
\affiliation{Nanotechnology Research Center,  Shahrekord University,  8818634141,  Shahrekord,  Iran}
\author{Sareh Shahidani}
\affiliation{Department of Physics, Sharif University of Technology, Tehran 14588, Iran}
\author{Somaye Mohamadi Abdhvand }
\affiliation{Department of Physics,  Faculty of Science,  Shahrekord University P. O. Box 115,  Shahrekord ,  Iran}

\date{\today}
\begin{abstract}
We study the geometric phase (GP) of a two-level atom coupled to an environment composed of free space and a dielectric nanosphere in thermal and out of thermal equilibrium. We analytically and numerically analyze the optical properties and loss of the dielectric medium, along with the non-equilibrium effects of the environment on the GP.  In the weak coupling limit, we find that the correction to the GP depends on the partial local density of photonic states at the atom position, and an effective parameter that emerges out of the non-equilibrium configuration of the system. The GP exhibits a significant enhancement due to the excitation of evanescent surface waves at its resonance frequency. It is shown that the GP acquired by the atomic system out of thermal equilibrium is always bounded between the thermal-equilibrium counterparts. Furthermore, the temperature difference between the nanosphere and free space can play an important role in the GP only at moderate atomic distances from the nanosphere. Our results elegantly demonstrate properties of the GP near material media that can support phononic modes and pave the way for further research of GP as a resource for quantum computation.
\end{abstract}
\maketitle
\section{Introduction}\label{intro}
Pancharatnam~\cite{Pancharatnam1975} was first to introduce the concept of
GP to the classical optics, and then Berry~\cite{Berry} discover this concept
in the quantum mechanics. He showed that a quantum system can acquire a GP upon
adiabatic transport of its Hamiltonian around a closed path in parameter space, or in the projective Hilbert
space. It depends only on the geometry of the path that is taken and occur in a wide range of circumstances in both classical and quantum systems, such as Foucault's pendulum~\cite{Berry1990} and Aharonov-Bohm effect~\cite{Aharonov1959}. Over the past decades, the original adiabatic GP has been generalized to different closed and open quantum systems undergoing nonadiabatic or noncyclic evolution~\cite{Aharonov,Anandan1,Anandan2,Samuel,Pati}. Recently, the GP has great potential for applications in quantum computation, quantum sensing and quantum information processing~\cite{KleiBler2018,Huang2019,Cho,Johnsson,Chen1,Zhang, Colmenar}.

The presence of decoherence and dissipation due to interactions with the environment can modify the GP. In such cases, the GP can still provide valuable information about the system-environment interaction and its effects on the system's evolution.
In the context of open quantum systems, different approaches have been applied to explore the modification of  the GP caused by the external environment~\cite{Whitney1,Whitney2,Carollo1,Carollo2,Tong, Marzlin,Lombardo,Villar1}. Understanding the influence of the environment on the GP is an important issue. It demonstrates the resilience of the GP to various types of noises, while also shedding light on how information about the environment is encoded in the GP. Many works have been conducted along these lines, considering different types of decoherence sources in both Markovian, and non-Markovian environments ~\cite{Lombardo2,Lombardo3,Lombardo4,Villar2,hu, Chen,Dechiara,Banerjee,Yi,Wang}, non-equilibrium environment~\cite{Cai2019}, and strong coupling regime~\cite{Villar2022}.

Two-level system, which is the focus of this study, and harmonic oscillator are the simplest and yet significant examples for exploring GP in the context of open quantum system. Measuring the GP of these simple systems can be used as a tool to get information about the environment~\cite{Yi,Wang,Fuentes-Guridi1,Fuentes-Guridi2} and detect quantum effects like the Unruh effect~\cite{Martin-Martinez1} and quantum friction force~\cite{Farias}.
In Ref. \cite{Fuentes-Guridi1}, GP encodes information about the number of particles in the surrounding quantized field. In Ref.~\cite{Yi}, it is shown that GP of a dephasing two-level system contains phase information of the environment. In Refs.~\cite{Martin-Martinez1,Martin-Martinez2}, it has been shown that the two-level system can be used as a high-precision quantum thermometer, when its GP is affected by surrounding quantized field in thermal state.

The purpose of this paper is to study how GP of a two-level system is affected by decoherence and dissipation of the environment composed of free space and a dielectric nanosphere.  Here, the dielectric medium with certain susceptibility and temperature modifies the quantum vacuum field. Depending on the relative distance between the atom and the dielectric medium as well as their geometrical and optical properties, this mechanism results in varying dissipation rate. Furthermore, when the atom is situated in a stationary configuration out of thermal equilibrium, where the medium temperature is maintained fixed and different from the surrounding free space, the density matrix of the atom evolves into a non-equilibrium steady state with an effective temperature~\cite{Bellomo1,Bellomo2}. Consequently, the GP of the atomic system can be expected to encode information about the geometrical and optical properties of the dielectric medium along with its temperature, and the non-equilibrium effects of the environment.

Recently, much attention has been paid to studying the non-equilibrium effects of the environment on dynamical evolution and decoherence of open quantum systems.
The lack of thermal equilibrium may appear in transient and ultrafast processes in physical or biological systems, providing significant possibilities not available in the thermal equilibrium state, such as the suppression of the decoherence and the non-Markovian features of the quantum systems~\cite{Cai2016}. In this study, we compute the GP of the atomic system under nonunitary dynamics in both equilibrium and non-equilibrium configurations. The environmentally induced corrections to the GP can be decomposed in different contributions: medium correction induced by the modified quantum vacuum field due to the presence of the nanosphere and corrections induced by out of thermal equilibrium effects of the system.

The structure of this paper is organized as follows. In Sec.~\ref{Model}, we introduce the model under investigation and obtain an exact master equation for the reduced atomic density operator to describe the dynamics of a two-level atom coupled to the quantized electromagnetic field in the presence of a nanosphere. We derive explicitly the atomic transition rates in terms of the partial local density of states and an effective parameter, which depend on material properties of the nanosphere, the atom position, and the orientation of the atom's dipole moment. In Sec.~\ref{Sec:Geometric phase}, we compute the GP acquired by
the atomic system in both equilibrium and non-equilibrium configurations.
Sec.~\ref{Num-sec} contains the numerical analysis of the medium- and non-equilibrium induced corrections to the GP when the nanosphere is made of the Gallium Arsenide, and compares them with those obtained when the atom is alone in free space. Finally, in Sec.~\ref{conclusion}, we make our final remarks and summery of the conclusions. Details on the derivation of the electromagnetic Green tensor of the system can be
found in Appendix~\ref{appendix}.

\section{The Model}\label{Model}
\begin{figure}[ht]
\centering
\includegraphics[width=0.7\columnwidth]{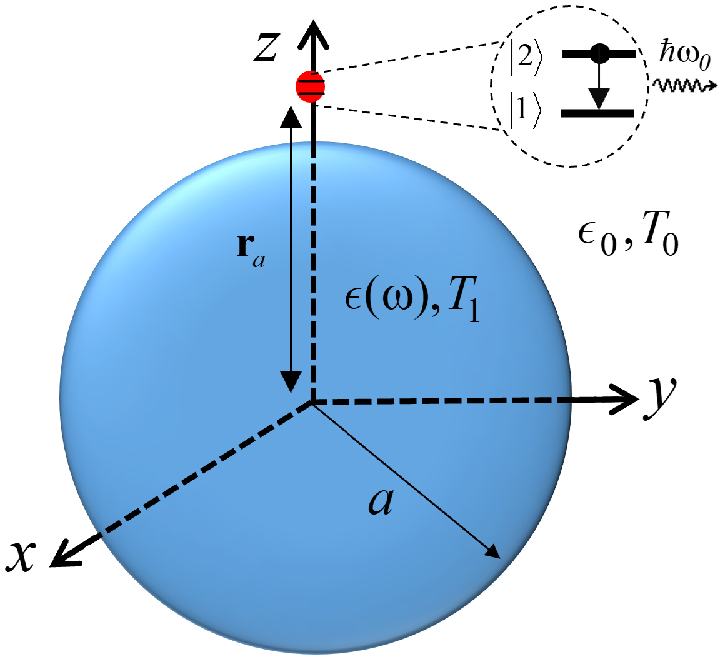}
\caption{(Color online) A schematic of the system that is out of thermal equilibrium but in a stationary regime. It consists of a two-level atom located at the distance ${\bf r}_{A}$ from the center of a dielectric sphere with permittivity function $\epsilon (\omega)$ and the radius $a$ at temperature $T_1$. The atom-nanosphere is embedded in a vacuum environment with free-space permittivity $\epsilon_0$ at temperature $T_0$.}
\label{fig:Nano_Sphere}
\end{figure}

We consider a two-level atom with the transition frequency $\omega_0$ in the vicinity of a dissipative-dispersive dielectric sphere with permittivity $\epsilon(\omega)$ and the radius $a$ in a stationary configuration out of thermal equilibrium, as schematically shown in Fig.~\ref{fig:Nano_Sphere}. The sphere and the surrounding vacuum are maintained at different constant temperatures, $T_1$ and $T_0$, respectively. The distance from the atom to the center of the nanosphere is ${\bf r}_a$.
The atom interacts with the sphere via the quantized electromagnetic field at the position of the atom.
The quantization of the electromagnetic field in the presence of dissipative and dispersive dielectric media is accomplished in a second-quantization framework\cite{Milonni}. The framework is well known and has been carried out through the canonical\cite{Huttner, Jeffers,Suttorp, Amooshahi, Kheirandish1,Philbin,Kheirandish2,Amooghorban,Gruner,Dung,Scheel,Matloob1,Knoll,Matloob2} and phenomenological approaches~\cite{Gruner,Dung,Scheel,Matloob1,Knoll,Matloob2}.
We follow the rigorous canonical approach where the dielectric medium is modeled as a reservoir consisting of infinite harmonic oscillators. These harmonic oscillators, which characterized by a
medium field, provide the polarizability and lossy characters of the medium. In this way, the dissipative medium enters directly into the quantization scheme, so that the complex frequency-dependent permeability of the medium is characterized through the macroscopic parameters of our model. Furthermore, the initial conditions of the medium field determines the explicit form of the current noise operator (more details can be found in Ref.~\cite{Kheirandish1,Behbahani,Amooghorban17}). Based on this quantization scheme, the Hamiltonian of the entire system is given by
\begin{equation}\label{H_t}
 {H} = {H}_{a}+  {H}_{f}+  {H}_{int},
\end{equation}
where
\begin{subequations}\label{H_}
\begin{eqnarray}
 {H}_{a}&=&\hbar {\omega _0} \sigma^{+}\sigma^{-},\\
 {H}_{f}&=& {\int {{d^3}{\bold{r}}} } \int\limits_0^\infty  {d\omega } \,\hbar \omega \,{\bold{ f}} ^\dag ({\bold{r}},\omega )\cdot {{\bold{ f}} }({\bold{r}},\omega  ),\\
 {H}_{int}&=&- \bold{D}\cdot \bold{E}({\bf r}_{a},t) \label{H_int}
\end{eqnarray}
\end{subequations}
are, respectively, the usual Hamiltonian of the free atom, the Hamiltonian of the medium-assisted electromagnetic
field, and the dipolar atom-field interaction Hamiltonian in the rotating wave approximation. Here,  $\sigma^{-}(\sigma^{+})$ denotes the lowering (raising) operator of the atom, the annihilation and creation operators ${\bold{ f}}$ and  ${\bold{ f}}^\dag$ represent the collective excitations of the electromagnetic fields and the medium and satisfy the canonical commutation relation $[ { f}_i({\bf r},\omega), { f}^{\dagger}_j({\bf r'},\omega') ]=\delta_{ij} \delta({\bf r}-{\bf r'})\delta(\omega-\omega')$, ${\bf E}({\bf r}_{a},t)$ is the electric field operator at the atom's position, and $ {\bold{D}}= \bold{d}_{21}\left| 2 \right\rangle \left\langle 1 \right|  +  \bold{d}_{21}^ * \left| 1 \right\rangle \left\langle 2 \right|$ is the electric dipole moment of the atomic system.
Given that the vacuum (dielectric sphere) is in local thermal equilibrium at constant temperature $T_0$ ($T_1$), the correlation functions between the elementary excitations of the system in a stationary configuration have the following form \cite{Wu}
\begin{subequations}\label{correlation func}
\begin{eqnarray}
\langle {\bf f}({\bf r},\omega){\bf f}^{\dagger}({\bf r'},\omega')\rangle&=&( 1+n(\omega,T_i))\nonumber\\&\times& \delta({\bf r}-{\bf r'})\delta(\omega-\omega'),\;\;\;\;\;\;\;\;\;\\
\langle {\bf f}^{\dagger}({\bf r},\omega){\bf f}({\bf r'},\omega')\rangle&=&n(\omega,T_i)\nonumber\\&\times&\delta({\bf r}-{\bf r'})\delta(\omega-\omega'),\;\;\;\;\;\;
\end{eqnarray}
\end{subequations}
where $n(\omega,T_i)=1/\big(\exp[\hbar\omega/k_B T_i]-1\big)$.

\subsection{Master equation}
In this paper, we intend to compute the GP of the atomic system interacting with the medium-assisted electromagnetic field during its evolution at stationary but nonequilibrium condition. In order to explore a general description of the
atom dynamics, we start from the von Neumann equation $\dot{\varrho}_{tot}(t)=-\frac{i}{\hbar}[H_{int}^I(t),\varrho_{tot}]$, where $\varrho_{tot}$ is the density operator of the combined close system, and $H_{int}^I$ is the interaction Hamiltonian~(\ref{H_int}) in the interaction representation.
Within the Born-Markov and rotating wave approximations, we trace out the photonic degrees
of freedom and obtain the following master equation of the atom alone whose projections on the
basis of the atomic subspace ~\cite{Breuer}
\begin{equation}\label{rho-dot}
\dot{\varrho} (t) = -\frac{i}{\hbar}\left[{H}_{a}+ {H}_{LS}, \varrho(t) \right]  + \mathcal{D}(\varrho (t) ).
\end{equation}
The first term in the above equation describes the unitary evolution of the
reduced density operator, and $\mathcal{D}( \varrho(t))$ is the Lindblad dissipator introducing dissipative effects such as
relaxation or dephasing. Here, the Lamb-shift Hamiltonian
\begin{equation}
{ H}_{LS} = \hbar \big( S(-\omega_0)|1\rangle \langle 1| +S(\omega_0)|2\rangle \langle 2| \big),
\end{equation}\label{Hls}
induces a shift of the atomic energy levels with a renormalized energy level spacing $\Omega(\omega_0)=\omega_0+S(\omega_0)-S(-\omega_0)$, wherein $S(\omega)=\Sigma_{i,j} s_{ij}(\omega) [{\bf d}_{21}]_{i}^* [{\bf d}_{21}]_{j}$ and $S(-\omega)=\Sigma_{i,j} s_{ij}(-\omega) [{\bf d}_{21}]_{i} [{\bf d}_{21}]_{j}^*$ with the parameter $s_{ij}$ defined by
\begin{eqnarray}\label{s_ij}
&&s_{ij}(\omega)=\frac{1}{\hbar^2}P\int_0^{\infty}d\omega_1\int_0^{\infty}d\omega_2\nonumber\\
&&\times \left[\frac{\langle E_i(\bold{r},\omega_2)E^{\dagger}_j(\bold{r'},\omega_1)\rangle}{\omega-\omega_2} +\frac{\langle E^{\dagger}_i(\bold{r},\omega_2)E_j(\bold{r'},\omega_1)\rangle}{\omega+\omega_2}\right],
\end{eqnarray}
where $P$ is the Cauchy principal value.
The Lindblad dissipator
\begin{eqnarray}\label{D}
\mathcal{D}( \varrho(t))&= &\Gamma(\omega) \big[ \varrho_{22}(t)|1 \rangle \langle 1 |  -  \frac{1}{2}\left\{ |2 \rangle \langle 2 |,\varrho (t)\right\} \big]\nonumber\\
&+&  \Gamma(-\omega) \big[ \varrho_{11}(t)|2 \rangle \langle 2 |  -  \frac{1}{2}\left\{ |1 \rangle \langle 1 |,\varrho (t)\right\} \big] ,
 \end{eqnarray}
contains the downward and upward atomic transitions $\Gamma(\omega)=\Sigma_{i,j} \gamma_{ij}(\omega) [{\bf d}_{21}]_{i}^* [{\bf d}_{21}]_{j}$ and $\Gamma(-\omega)=\Sigma_{i,j} \gamma_{ij}(-\omega) [{\bf d}_{21}]_{i} [{\bf d}_{21}]_{j}^*$, where the rates $\gamma_{ij}(\pm \omega)$ are expressed in terms of the Fourier transform of the electric field correlation functions as follows~\cite{Hu,Wu}
\begin{eqnarray}\label{gamma_ij}
\gamma_{ij}(\omega)=\frac{2\pi}{\hbar^2}\int_0^{\infty}d\omega'\times\left\{\begin{array}{*{20}{c}}
 \langle E_i(\bold{r},\omega)E^{\dagger}_j(\bold{r},\omega')\rangle,\;&&\omega>0\\
\langle E^{\dagger}_i(\bold{r},-\omega)E_j(\bold{r},\omega')\rangle,\;&&\omega<0
\end{array} \right. \nonumber\\
 \end{eqnarray}
%

\subsection{Transition rates out of thermal equilibrium}
Based on the canonical quantization scheme, the frequency component of the
electric field is given by
\begin{equation}\label{E}
\bold{E}({\bf r},\omega)=\frac{i\omega^2}{c^2}\sqrt{\frac{\hbar}{\pi\epsilon_0}}\int d^3{\bf r'}\sqrt{\mathrm {Im}\epsilon({\bf r'},\omega)} \bar{\bar {\bf G}} ({\bf r},{\bf r'},\omega)\cdot {\bf f}({\bf r'},\omega),\;
\end{equation}
where ${\rm Im}$ stands for the imaginary part, and $\bar{\bar{\bf G}}$ is the electromagnetic Green tensor fulfilling the Helmholtz equation $\big[ {\boldsymbol\nabla} \times{\boldsymbol\nabla} \times\, -\epsilon(\omega)\omega^2/c^2\big]\bar{\bar{\bf G}}({\bf r},{\bf r}')={\bar{\bf I}}\delta({\bf r}-{\bf r}')$ together with appropriate boundary conditions. Here, ${\bar{\bf I}}$ is the unit dyadic.

Using Eqs.~(\ref{correlation func}),~(\ref{gamma_ij}) and~(\ref{E}), and the Green identity $\omega^2\int   d^3 {\bf s} \,  {\rm Im} \epsilon({\bf s},\omega) \, {\bar{\bar {\bf G}}} ({\bf r},{\bf s},\omega) \cdot {\bar{\bar {\bf G}}}^{*}({\bf s},{\bf r'},\omega)= {c^2 } {\rm Im} {\bar{\bar {\bf G}}}({\bf r},{\bf r'},\omega)$, the downward and upward atomic transitions $\Gamma (\omega) $ and $\Gamma (-\omega) $ are written as~\cite{Wu}
\begin{eqnarray}\label{Gamma_}
\left( \begin{array}{*{20}{c}}\Gamma(\omega_0)\\\Gamma(-\omega_0)
\end{array}\right) =\Gamma_0\frac{\rho(\hat{{\bf n}}_d,{\bf r}_a,\omega_0)}{\rho_0}  \left(  \begin{array}{*{20}{c}}1+n_{eff}(\omega_0)\\n_{eff}(\omega_0)\end{array}\right)\,\,\,
\end{eqnarray}
where $\Gamma_0=\frac{\omega_0^3\mid{\bf d}_{12}\mid^2}{3\pi\epsilon_0\hbar c^3}$ is the vacuum spontaneous emission rate. Here, the partial local density of states(PLDOS) is defined as~\cite{Novotny2006}
\begin{eqnarray}\label{alpha}
\rho(\hat{{\bf n}}_d,{\bf r},\omega)&=&\frac{6\omega}{\pi c^2} \big( \hat{{\bf n}}_d\cdot \mathrm{Im}{\bar{\bar{\bf G}}}({{\bf r}},{{\bf r}},\omega)\cdot \hat{{\bf n}}_d^* \big),
\end{eqnarray}
where ${\bf n}_{d}$ is a unit vector pointing in the direction of the dipole moment ${\bold{d}}_{21}$. This quantity, which depends on material properties of the sphere, the atom position, and the orientation of the atom's dipole moment, measures the number of photonic states per unit of frequency and volume at a certain position and frequency due to the presence of the sphere. In particular, in free space, after averaging Eq.~(\ref{alpha}) over different dipole orientation, $\rho$ is simply $\rho_0={\omega_0^2}/{\pi^2 c^3}$. The effective parameter
\begin{eqnarray}\label{neff}
n_{eff}(\omega_0)&=&n(\omega_0,T_0)\\
&+&\frac{\rho_m(\hat{{\bf n}}_d,{\bf r}_a,\omega_0)}{\rho(\hat{{\bf n}}_d,{\bf r}_a,\omega_0)}\big(n(\omega_0,T_1)-n(\omega_0,T_0)\big),\nonumber
\end{eqnarray}
depends both on temperatures and the material properties of the sphere. This material dependence, emerging out of the non-equilibrium configuration of the system, is established through the quantity:
\begin{eqnarray}\label{g_ij}
\rho_m(\hat{{\bf n}}_d,{\bf r}_a,\omega)&=&\frac{6\omega^3}{\pi c^4} \int_0^a d r'  r'^2 \mathrm{ Im}\:\epsilon(r')\int  d\phi' d\theta' \sin\theta' \nonumber\\
&\times&  \big( \hat{{\bf n}}_{d}\cdot {\bar{\bar{\bf G}}}({\bf r}_a,{\bf r'},\omega) \cdot {\bar{\bar{\bf G}}}^*({\bf r'},{\bf r}_a,\omega)\cdot \hat{{\bf n}}_{d}^* \big),\,\,\,\,\,\,\,\,\,\,
\end{eqnarray}
which from now on we call the medium PLDOS.
Likewise, one can define its vacuum counterpart, i.e., $\rho_{v}(\hat{{\bf n}}_d,{\bf r}_a,\omega)$, where the lower and upper bounds of the above integral are replaced by $a$ and $\infty$, respectively. These two quantities lead to the PLDOS, $\rho=\rho_m+\rho_{v}$, through the Green identity. Note that $\rho_m$ and $\rho_{v}$ are positive quantities since $\mathrm{ Im}\:\epsilon\geq 0$ and their integrals can be recast into a form $ \int |g|^2$ using the reciprocity theorem, ${\bar{\bar{ G}}}_{ij}({\bf r'},{\bf r}_a,\omega)={\bar{\bar{ G}}}_{ji}({\bf r}_a,{\bf r}',\omega)$. With this in mind, the effective parameter~(\ref{neff}) can be rearrange to give $n_{eff}=\big(n(\omega,T_0)\rho_{v}+n(\omega,T_1)\rho_{m}\big)/\rho$, resulting in the inequality $n(\omega,T_{min})\leq n_{eff} \leq n(\omega,T_{max})$ where $T_{min}=min(T_0,T_1)$ and $T_{max}=max(T_0,T_1)$. Subsequently, the downward and upward atomic transitions~(\ref{Gamma_}) are bounded between their equilibrium counterpart values at temperatures $T_0$ and $T_1$. These findings are in agreement with the results reported in~\cite{Bellomo2}.

In the case $T_1=T_0=T$, the sphere is in the thermal equilibrium with the background thermal radiations and $n_{eff}(\omega_0)$ reduces to the mean thermal photon number $n(\omega_0,T)$. Therefore, the atomic transitions $\Gamma(\pm \omega_0)$ are expressed as a product of $\Gamma_0 \rho/\rho_0$ with the factor $1+n(\omega_0,T)$ or $n(\omega_0,T)$ at the thermal equilibrium, as expected.

Given these results, the dynamic evolution of the elements of the reduced atomic density operator~(\ref{rho-dot}) can be written as
\begin{subequations}\label{Differential equation density operator}
\begin{eqnarray}
\dot{\varrho}_{11}(t) &=&-\Gamma (-\omega _0)\varrho_{11}(t) + \Gamma (\omega _0)\varrho_{22}(t),\\
\dot{\varrho}_{22}(t) &=& \Gamma (-\omega _0)\varrho_{11}(t) - \Gamma (\omega _0)\varrho_{22}(t),\\
\dot{\varrho}_{12}(t) &=& \left[ i(\Lambda +\omega_0) -  \frac{\Gamma (\omega _0)+\Gamma (-\omega _0)}{2}\right]\varrho_{12}(t),\,\,\,
\end{eqnarray}
\end{subequations}
where $\Lambda\equiv \Omega-\omega_0$ is the so-called Lamb shift. Using Eqs.~(\ref{s_ij}) and~(\ref{E}) and the Kramers-Kronig relations the Lamb shift is given by
\begin{eqnarray}
\Lambda(\omega_0)= -\frac{\omega}{\hbar\epsilon_0 c^2}\big( \hat{{\bf n}}_{d}\cdot \mathrm{Re}{\bar{\bar{\bf G}}}({{\bf r}_a},{{\bf r}_a},\omega_0)\cdot \hat{{\bf n}}_{d}^* \big),
\end{eqnarray}
which is independent of the temperatures. This implies that the Lamb shift does not depend on the absence or
 presence of thermal equilibrium.

\section{Geometric phase}\label{Sec:Geometric phase}
In this section, we turn to compute the GP associated with the evolution of the two-level atom in a stationary configuration out of thermal equilibrium.
It is well known that the atom evolves nonunitarily due to its coupling with the medium-assisted electromagnetic field near the nanosphere whose temperature is kept fixed and different from that of the surrounding vacuum. We pursue the kinematic approach of Tong et al.~\cite{Tong}, which gives the GP for a quantum system under nonunitary dynamical evolution, and it reads
%
\begin{equation}\label{phi_def}
\Phi=\arg \sum_{k}\sqrt {\varepsilon_k(0 )\varepsilon_k(t)}\langle\psi_k(0)|\psi_k(t)\rangle
e^ {-\int_0^{t}dt'\langle\psi_k(t')|\dot{\psi}_k(t')\rangle},\;\:\:\:
\end{equation}
where $\varepsilon_k(t)$  and $|\psi_k(t)\rangle$ are the eigenvalues and eigenvectors
of the reduced density matrix $\varrho(t)$. As can be seen from the definition above, to compute the GP, it is important to first find the solution of time-dependent reduced density matrix~(\ref{Differential equation density operator}) at all times.
To this end, we assume that the system is initially in the pure state
\begin{equation}\label{initial_state}
\left|\psi(0) \right\rangle=\cos\dfrac{\theta_0}{2}\left|2 \right\rangle +\sin\dfrac{\theta_0}{2}\left|1 \right\rangle,
\end{equation}
Using Eqs.~(\ref{Differential equation density operator}), the elements of reduced density matrix for times $t>0$ are given by
\begin{subequations}
\begin{eqnarray}
\varrho_{11}(t)&=&\sin^2\frac{\theta_0}{2}e^{-\Gamma_{+} t}+\frac{\Gamma_{+}+\Gamma_{-}}{2\Gamma_{+}}(1-e^{-\Gamma_{+} t}),\,\,\,\, \\
\varrho_{22}(t)&=&\cos^2\frac{\theta_0}{2}e^{-\Gamma_{+} t}+\frac{\Gamma_{+}-\Gamma_{-}}{2\Gamma_{+}}(1-e^{-\Gamma_{+} t}),\,\,\,\,\, \\
\varrho_{12}(t)&=&\varrho_{21}^*(t)=\dfrac{1}{2}e^{( i\Omega-\Gamma_+/2) t} \sin\theta_0,
\end{eqnarray}
\end{subequations}
where $\Gamma_{\pm}\equiv\Gamma(\omega_0)\pm\Gamma(-\omega_0)$, with $\Gamma_{+}$ being the dephasing decay rate of the quantum coherence of the system.
Solving the eigenvalue problem for the reduced density matrix, the eigenvalues of $\varrho(t)$ are obtained as
\begin{eqnarray}\label{eigenvalues}
&&\varepsilon_{\pm}(t)=\frac{1}{2}\\
&&\left(1\pm \sqrt{\sin^2\theta_0 e^{-\Gamma_+t}+(\cos\theta_0 e^{-\Gamma_+t}-Q(1-e^{-\Gamma_+t}))^2}\right),\nonumber
\end{eqnarray}
with $Q\equiv\Gamma_-/\Gamma_+$. The corresponding eigenvectors of $\varrho(t)$ can be expressed as
\begin{eqnarray}\label{eigenvectors}
|\psi_{+}(t)\rangle&=&e^{i\Omega t}\cos\frac{\theta_t}{2}|1\rangle+\sin\frac{\theta_t}{2}|2\rangle,\\
|\psi_{-}(t)\rangle&=&e^{i\Omega t}\sin\frac{\theta_t}{2}|1\rangle-\cos\frac{\theta_t}{2}|2\rangle,
\end{eqnarray}
where $\tan (\theta_t/2)=\sqrt{(\varrho_{22}-\varepsilon_-)/(\varrho_{11}-\varepsilon_-)}$. Since $\varepsilon_{-}(0)=0$, we only need $|\psi_+(t)\rangle$ to calculate the GP. Substituting Eqs.~(\ref{eigenvalues}) and~(\ref{eigenvectors}) into Eq.~(\ref{phi_def}), the GP reduces to
%
%
\begin{equation}\label{phi}
\Phi=\arg {\langle\psi_k(0)|\psi_k(t)\rangle} -\Omega\int_0^{t} dt' \cos^2(\theta_{t'}/2).
\end{equation}
here, the first term is the Pancharatnam phase~\cite{Pancharatnam1975}, which results from the correlation
of the time-evolved state $|\psi_k(t)\rangle$ and the initial state $|\psi_k(0)\rangle$, is written by
\begin{equation}\label{Pancharatnam phase}
\arg {\langle\psi_k(0)|\psi_k(t)\rangle}=\frac{\sin \Omega t}{\cos \Omega t+\tan (\theta_t/2)\coth (\theta_t/2)}.
\end{equation}
For the case that the system evolves along a quasicyclic path with the evolution time $t = 2\pi/\Omega$, the Pancharatnam phase~(\ref{Pancharatnam phase}) vanishes and makes no contribution to the GP.
The last term in Eq.~(\ref{phi}) arising from the GP of
the dynamical evolution, is given by
\begin{eqnarray}\label{phi_t}
\Phi&=& -\frac{\Omega}{2}\int_0^T dt\\
&&\times (1-\frac{Q-Q e^{\Gamma_+ t}+\cos\theta_0}{\sqrt{\sin^2\theta_0 e^{\Gamma_+ t}+(Q-Q e^{\Gamma_+ t}+\cos\theta_0)^2 }}).\nonumber
\end{eqnarray}
A direct calculation of this integral is rather tedious. However, a simple estimate can reveal that the ratio $\Gamma_0/\omega_0$ is vanishingly small. For instance, considering a free space qubit with the transition frequency $\omega_0/2\pi = 4.68 {\rm GHz}$ and the relaxation time $2.65 \mu s$~\cite{Leek2007}, $\Gamma_0/\omega_0$ is of order $10^{-5}$. This allows us to evaluate this integral by making a series expansion of the integrand in terms of $\Gamma_0/\omega_0$. Up to leading order in $\Gamma_0/\omega_0$, it leads to
\begin{eqnarray}\label{phi_approx1}
\Phi\approx&&-\pi(1-\cos\theta_0)\\
&&-\frac{\Omega }{2\omega_0}\Gamma_0 \sin^2\theta_0 \int_0^T dt\, t (2Q+\cos\theta_0) \frac{\partial \Gamma+}{\partial(\Gamma_0/\omega_0)},\nonumber
\end{eqnarray}
The first term of Eq.~(\ref{phi_approx1}) is the GP for the isolated system with no influence from the environment, and the second term is the correction originates from the atom-electromagnetic field interaction in the absence of thermal equilibrium. Using the definition~(\ref{Gamma_}) for $\Gamma(\pm\omega_0)$, the geometric phase~(\ref{phi_approx1}) can be rearranged to give:
\begin{eqnarray}\label{phi_approx2}
\Phi \approx-\pi(1-\cos\theta_0)&-&\frac{\pi^2 \Gamma_0}{2\rho_0 \omega_0 }\rho(\hat{{\bf n}}_d,{\bf r}_a,\omega_0)\sin^2\theta_0\\&\times &\left[ \cos\theta_0+2 n_{eff}(\omega_0)\cos\theta_0+2\right]\nonumber ,\,\,\,\,
\end{eqnarray}
As can be seen, in the absence of the electromagnetic modes of the free space and the
background thermal radiations, we can recover the familiar expression $\Phi_0 \approx-\pi(1-\cos\theta_0)$ for the GP~\cite{Carollo1, Marzlin, Chen}. In what follows, we separate the contribution induced by the non-equilibrium effects in the presence of dielectric nanosphere from that of $\Phi_0$ as
\begin{eqnarray}\label{phi_approx3}
\Delta\Phi&=&\frac{\pi^2 \Gamma_0}{2\rho_0 \omega_0 }\rho(\hat{{\bf n}}_d,{\bf r}_a,\omega_0)\sin^2\theta_0\\
& &\times\left[ \cos\theta_0+2 n_{eff}(\omega_0)\cos\theta_0+2\right]\nonumber ,\,\,\,\,
\end{eqnarray}
where $\Delta\Phi =\vert\Phi -\Phi _0\vert$. This GP difference reflects the correction to the unitary GP and becomes significant for nanostructures supporting plasmonic or phononic modes due to the increase of the PLDOS close to the nanosphere at the resonance frequency, as will be seen in the next section.
%

\section{NUMERICAL ANALYSIS}\label{Num-sec}
In this section, we present a numerical analysis of the GP~(\ref{phi_approx3}) for the case where the atom is located on the z-axis at a distance of $r_a=1.7 \mu m$ from the center of a nanosphere with the radius of $a=700 nm$ for both out of thermal equilibrium and in thermal equilibrium configurations. We consider the dipole moment of the atom oriented perpendicularly to the surface of the GaAs nanosphere, \textrm{i.e.,} $\hat{\bf n}_d=\hat{\bf r}$. Similar results are obtained for the case where the dipole moment of the atom is tangent to the surface of the nanosphere (not shown here). Here, the Gallium Arsenide (GaAs) nanosphere is investigated, which supports surface waves known as localized surface phonon-polaritons(LSPP), with optical properties well described by the Drude-Lorentz model,
\begin{equation}\label{Drude}
\epsilon(\omega)=\epsilon_{\infty}\frac{\omega^2-\omega_l^2+i\gamma_e\omega}{\omega^2-\omega_r^2+i\gamma_e\omega},
\end{equation}
%
%
%
where $\epsilon_{\infty}=11$, $\gamma_e=0.00452\times10^{14} (rad  /s)$,  $\omega_r=0.506\times 10^{14} (rad  /s)$, and  $\omega_l=0.550\times 10^{14}(rad  /s)$~\cite{Palik}. The GaAs nanosphere embedded in vacuum exhibits a dipolar surface-phonon mode at a frequency dictated by the condition ${\rm Re}[\epsilon(\omega)]=-2$~\cite{Kittle1987}. Using the Drude-Lorentz model~(\ref{Drude}), this resonance appears at the frequency of $1.074 \omega_r$.
In what follows, we assume that the atomic frequencies are of the same order as the resonance frequency of the sphere. Such frequencies can be achievable using artificial atoms made of semiconductor quantum dots~\cite{Zibik}.
%

\subsection{Medium-induced corrections to the GP difference}
\begin{figure}[ht]
\centering
\includegraphics[width=\columnwidth]{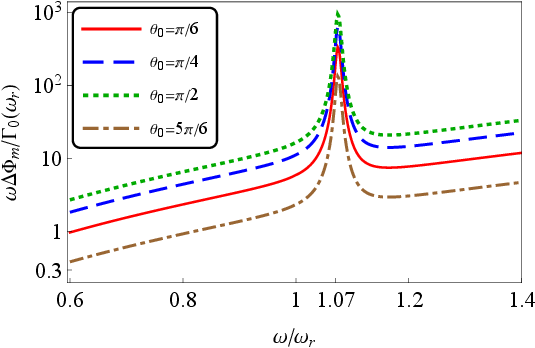}
\caption{(Color online). The normalized medium-induced GP $\omega\Delta\Phi_m/\Gamma_0(\omega_r)$, vs. normalized frequency $\omega/\omega_r$ for four initial states of $\theta_0=\pi/6$ (red solid line), $\theta_0=\pi/4$ (blue dashed line), $\theta_0=\pi/2$ (green dotted line) and $\theta_0=5\pi/6$ (brown dot-dashed line). The nanosphere is made of GaAs with the permittivity function~(\ref{Drude}). The atom is located on the z-axis at a distance of ${ r}_{a}=1.7 \mu m$ from the center of the GaAs nanosphere with the radius of $a=700 nm$.}
\label{fig:phiv1}
\end{figure}

In this subsection, we restrict our attention to the GP difference alone due to the presence of the nanosphere. To this end, we consider that the whole system is in thermal equilibrium at zero temperature. For this particular case of zero temperature, due to Eq.~(\ref{neff}), $n_{eff}$ vanishes and the medium-induced correction to the GP~(\ref{phi_approx3}) is obtained as
\begin{equation}\label{Phi_m}
\Delta\Phi_m=\frac{\pi^2 \Gamma_0}{2\rho_0\omega_0 }\rho(\hat{\bf r},{\bf r}_a,\omega_0)\sin^2\theta_0\left[ \cos\theta_0+2\right],
\end{equation}
which depends on the properties of the atom via the transition frequency $\omega_0$ and the spontaneous emission in vacuum $\Gamma_0$, the initial state $\theta_0$, and material and geometrical properties of the nanosphere through $\rho$.
In the absence of the dielectric sphere, where $\rho=\rho_0$, the GP difference~(\ref{Phi_m}) reduces to $\pi^2 \Gamma_0\sin^2\theta_0\left[ 2\cos\theta_0+1\right]/\omega_0$, which is just the GP difference acquired by a two-level
atom in vacuum. This is consistent with the result obtained for a two-level atom coupled to an environment with Lorentzian spectral density~\cite{Chen}.
However, as depicted in Fig.~\ref{fig:phiv1}, the presence of the sphere disrupts this result due to supporting surface-phonon modes.
This figure illustrates the normalized medium-induced GP, $\omega\Delta\Phi_m/\Gamma_0(\omega_r)$,  as a function of normalized frequency $\omega/\omega_r$ for different initial states of $\theta_0$. The GP difference, which is proportional to the PLDOS, shows a peak at $\omega=1.074\omega_r$ that corresponds to the LSPP resonance. This increases the PLDOS close to the sphere, which leads to a significant GP difference and provides ideal conditions for GP detection. It further shows that the GP difference enhances as $\theta_0$ increases from zero to $\pi/2$, while it decreases as $\theta_0$ increases beyond $\pi/2$.
This is understandable because the maximum of $\Delta\Phi_m$ occurs around $\theta_0=\pi/2$ with the maximum medium-induced GP $\pi^2\Gamma_0\rho(\omega_0)/\rho_0\omega_0$ where $\Phi_0$ also reaches its maximum, whereas $\Delta\Phi_m$ vanishes when $\theta_0=0$ and $\theta_0=\pi$.

\begin{figure}[ht]
\centering
\includegraphics[width=\columnwidth]{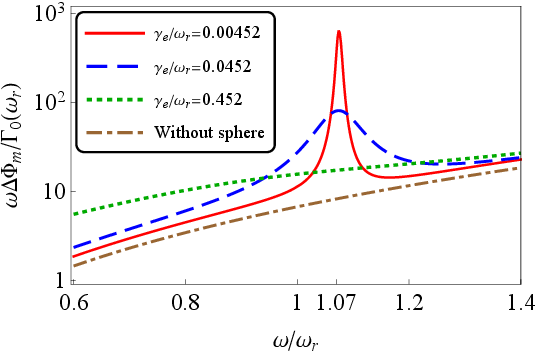}
\caption{(Color online). The normalized medium-induced GP $\omega\Delta\Phi_m/\Gamma_0(\omega_r)$, vs. the normalized frequency $\omega/\omega_r$ for three values of the damping coefficient $\gamma_e=0.00452\omega_r$ (red solid line), $\gamma_e=0.0452\omega_r$ (blue dashed line), and $\gamma_e=0.452\omega_r$ (green dotted line). As a reference, the GP difference in the absence of the nanosphere is plotted by the brown dot-dashed line. Here, $\theta_0=\pi/4$, and other parameters are the same as those in Fig.~\ref{fig:phiv1}.}
\label{fig:phiv2}
\end{figure}
Fig.~\ref{fig:phiv2} shows that the damping coefficient $\gamma_e$ can significantly influence on the medium-induced GP difference. The GP difference is always larger than that in the absence of the nanosphere at given frequencies, so that it is more pronounced in the LSPP resonance. Furthermore, the peak-width become broader with increasing $\gamma_e$ because the linewidth of the PLDOS is closely associated with the damping coefficient. In this sense, the dissipative effect of the sphere has a strong correction around the LSPP frequency. Far from the LSPP resonance, the GP difference is relatively small, but it increases slightly for larger values of $\gamma_e$. Since the PLDOS contains two radiative and nonradiative channels through which the atom can decay, therefore, this enhancement can be attributed to the increase in the nonradiative decay to the absorption modes in the nanosphere. Interestingly, our result indicates the
insensitivity of the GP to the presence of the nanosphere far from (near) the LSPP resonance frequency when the absorption is weak (strong). This means that the GP difference shows a stronger resilience to the medium with a large $\gamma_e$ near the LSPP resonance frequency.

%
\begin{figure}[ht]
\centering
\includegraphics[width=\columnwidth]{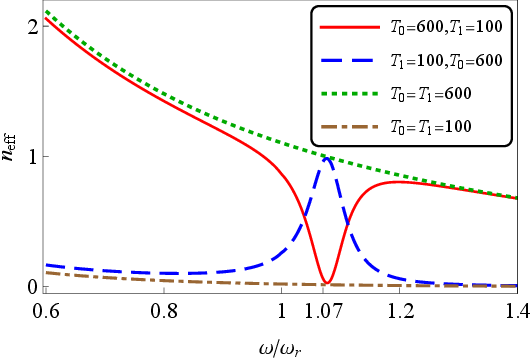}
\caption{(Color online) The effective parameter $n_{eff}$, vs. the normalized frequency $\omega/\omega_r$. Two configurations out of thermal equilibrium with temperatures $T_0=600K, T_1=100K$ (red solid line) and $T_0=100K, T_1=600K$ (blue dashed line) are compared with thermal equilibrium configurations at temperatures $T_0=T_1=600K$ (green dotted line) and $T_0=T_1=100K$ (brown dot-dashed line). The parameters we chose are the same as those in Fig.~\ref{fig:phiv1}.}
\label{fig:non-equ_1}
\end{figure}

\begin{figure}[ht]
\centering
\includegraphics[width=\columnwidth]{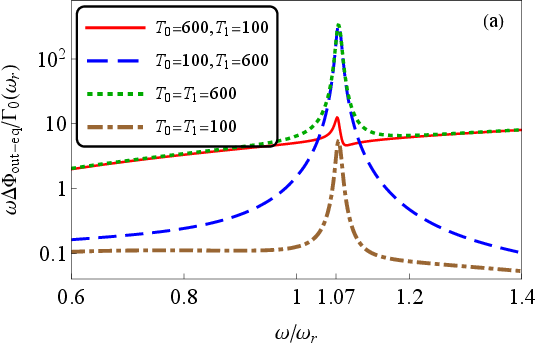}
\includegraphics[width=\columnwidth]{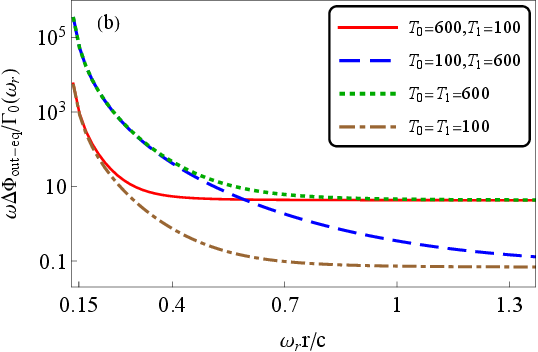}
\caption{(Color online) The normalized GP deviation $\omega \Delta\Phi_{out\text{-}eq}/\Gamma_0(\omega_r)$, vs. the normalized frequency $\omega/\omega_r$ (a) and  the normalized distance $\omega_r r/c$ (b), for both in thermal equilibrium and out of thermal equilibrium configurations. In (a), $r_a=1.7 \mu m=0.29 c/\omega_r$, and in (b) $\omega=1.074\omega_r$. The other parameters are the same as in Figs.~\ref{fig:phiv1} and~\ref{fig:non-equ_1}. }
\label{fig:non-equ_2}
\end{figure}

\subsection{Out of thermal equilibrium induced corrections to the GP in the presence of the nanosphere}
In the following, we mainly concentrate on out of thermal equilibrium corrections to the GP defined as
\begin{equation}\label{phi_th}
\Delta\Phi_{out\text{-}eq}=\frac{\pi^2\Gamma_0}{\rho_0\omega_0}\rho(\hat{\bf r},{\bf r}_a,\omega_0)n_{eff}(\omega_0)\sin^2\theta_0 \cos\theta_0,
\end{equation}
where the material property of the nanosphere is encoded on both quantities $\rho$ and $n_{eff}$. Likewise, the thermal equilibrium corrections to the GP, \textrm{i.e.}, $\Delta\Phi_{eq}$, can be obtained by replacing $n_{eff}$ in the above equation with the mean thermal photon number $n$.
Fig.~\ref{fig:non-equ_1} indicates the temperature and frequency dependence of $n_{eff}$, appearing in~(\ref{phi_th}), for two configurations out of thermal equilibrium at $T_0=600K, T_1=100K$ and $T_0=100K, T_1=600K$, and compares them with the corresponding thermal-equilibrium parameter, \textrm{i.e.,}  $n$, at temperatures $T_0=T_1=T_{max}=600K$ and $T_0=T_1=T_{min}=100K$. It is observed that  $n_{eff}$ is enclosed between $n(\omega,T_{min})$ and $n(\omega,T_{max})$, which is also established in Sec.~(\ref{Model}), and shows a significant variations around the LSPP resonance.
As a consequence, the GP difference $\Delta\Phi_{out\text{-}eq}$, which is proportional to $n_{eff}$, is always bounded between the thermal-equilibrium counterparts $\Delta\Phi_{eq}(T_{min})$ and $\Delta\Phi_{eq}(T_{max})$, as shown in Figs.~\ref{fig:non-equ_2}.
In Fig.~\ref{fig:non-equ_2} (a), the GP difference as a function of the frequency follows similar variations both in thermal equilibrium and out of thermal equilibrium, showing a strong peak around the LSPP resonance. However, the peak height is almost small when the nanosphere is at a lower temperature than the vacuum environment, although it is still larger than the thermal-equilibrium configuration at the sphere temperature (compare red solid and brown dot-dashed lines in Fig.~\ref{fig:non-equ_2} (a)). This is because, for the small atom-sphere separations considered here, the temperature of the sphere plays a significant role in the GP difference at the resonance frequency. It is evident from Fig.~\ref{fig:non-equ_2} (b) that for small values of the dimensionless distance $\omega_r r/c$ only the temperature of the nanosphere contributes, while at intermediate atomic distances both sphere and vacuum temperatures play a role in the GP difference. In contrast, for large values of $\omega_r r/c$, the out of thermal equilibrium induced GP tends towards the equilibrium counterpart at the vacuum temperature. This means that at large distances, only the vacuum temperature affects the GP difference. Therefore, at large (small) atomic distances, the GP difference at the LSPP resonance can vary with the temperature of the vacuum (sphere).

\subsection{Total corrections to the GP in the presence of the nanosphere}
\begin{figure}[ht]
\centering
 	\includegraphics[width=\columnwidth]{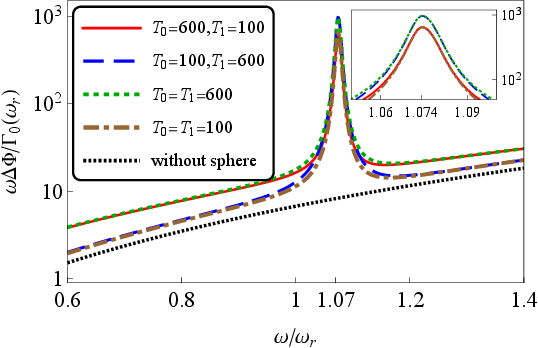}
\caption{(Color online) The normalized total GP deviation $\omega \Delta\Phi/\Gamma_0(\omega_r)$ as a function of the normalized frequency $\omega/\omega_r$ for both in thermal equilibrium and out of thermal equilibrium configurations. The black dotted line shows the GP difference in the absence of the nanosphere at the vacuum temperature of $100K$. Here, $\theta_0=\pi/4$, and other parameters are the same as in Figs.~\ref{fig:phiv1} and~\ref{fig:non-equ_1}. The inset shows the zoomed region near the LSPP resonance.}
\label{fig:non-equ_3}
 \end{figure}

In this subsection, we consider the combination of both medium and out of thermal equilibrium corrections to the GP in the presence of the nanosphere.
Fig.~\ref{fig:non-equ_3} shows the sum of these corrections, {\textrm{i.e.,}} Eq.~(\ref{phi_approx3}), as a function of the dimensionless frequency $\omega/\omega_r$.  We compare the GP difference at thermal-equilibrium at $T_0=T_1=100K$ and $T_0=T_1=600K$ with the GP out of thermal equilibrium at $T_0=600K, T_1=100K$ and $T_0=100K, T_1=600K$. As a reference, the GP difference is also plotted in the absence of the nanosphere at the vacuum environment temperature of $100K$ (the black dotted line).
As can be seen, the GP deviation $\Delta\Phi$ shows a considerable increase at the LSPP resonance for both in thermal and out of thermal equilibrium. However, regardless of the temperature of the sphere and the environment, there is a slight difference between these curves for the atom distance considered here. This is due to the prominent role of the medium-induced correction to the GP at the LSPP frequency and demonstrates almost the robustness of the GP to the temperature difference of the sphere and the environment. Far from the LSPP resonance, the GP difference decreases sharply for out of thermal equilibrium configurations and approaches the equilibrium counterparts at the vacuum environmental temperatures. In this sense, a small variation of frequency around the LSPP resonance lead to drastically modifications of the GP difference. This can provide evidence of thermally excited surface evanescent waves.
 \begin{figure}[ht]
\centering
 	\includegraphics[width=\columnwidth]{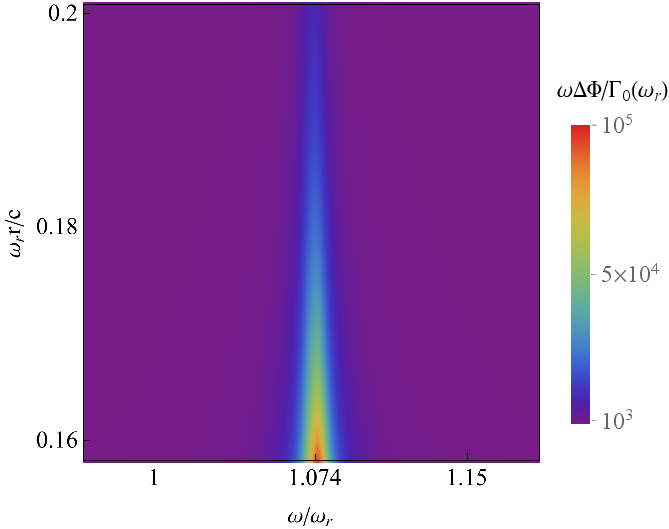}
\caption{(Color online) Density plot of the GP deviation $\Delta\Phi$ as functions of the normalized distance $\omega_r r/c$ and the normalized frequency $\omega/\omega_r$ for out of thermal equilibrium configuration with temperature $T_0=600K, T_1=100K$. Here, $\theta_0=\pi/4$, and the other parameters are the same as in Fig.~\ref{fig:phiv1}.}
\label{fig:non-equ_4}
 \end{figure}
In Fig.~\ref{fig:non-equ_4}, the GP difference $\Delta\Phi$ is shown as a function of the dimensionless frequency $\omega/\omega_r$ and the dimensionless distance $\omega_r r/c$ for
$T_0=600K$ and $ T_1=100K$. This illustrates the role of LSPP resonance in the GP difference. In particular, it exhibits a peak centered at the LSPP resonance for any value of the atom distance. Close to the nanosphere,
the PLDOS increases in the presence (absence) of evanescent surface waves confined near the surface of the sphere and strongly enhances (decreases) the GP difference.
Far from the nanosphere, the PLDOS decreases due to the absence of evanescent radiation, resulting in a decrease in the GP difference.
%

\section{Conclusion}\label{conclusion}
We have studied the corrections to the GP under a nonunitary evolution induced by the presence of a dielectric nanosphere.
We have considered the atom to be embedded in a stationary configuration out of thermal equilibrium, where the medium temperature is kept fixed and different from the surrounding free space. The effect of optical and geometrical properties of the dielectric medium along with its temperature on the geometric phase has been explored analytically and numerically. It was demonstrated that the first-order correction to the GP, which is proportional to the partial local density of states and the effective parameter $n_{eff}$, is significantly large close to the nanosphere when the surface phonon modes excited.
In this sense, a small change in frequency around the resonance frequency leads to drastic changes in the GP, providing ideal conditions for the GP detection.

On the other hand, the GP has shown a stronger resilience to the medium with a large damping coefficient near the localized surface phonon-polaritons resonance frequency.
For the small atom-sphere separations, the medium temperature leaves its footprint in the GP acquired by the atom, while for large distances, the sphere temperature is immaterial, and the out-of-thermal equilibrium-induced GP approaches the equilibrium counterpart at the free space temperature.
This suggest that the geometric phase can be used to construct a quantum thermometer for dielectric media in non-equilibrium regime.
Our results beautifully demonstrate properties of the GP near material media that support phononic or even plasmonic modes both in thermal and out of thermal equilibrium configurations and serve as a stepping stone for further research of GP as a resource for quantum computation or quantum sensing.

\appendix
\include{app1}
%
\section{ Green Tensor of the system}\label{appendix}
Following the method of scattering superposition in Refs.~\cite{Behbahani,Amooghorban17,Li1994}, we can write the electromagnetic Green tensor of the nanosphere in Fig.~\ref{fig:Nano_Sphere} in the form ${\bar{\bar{\bf G}}}({\bf r},{\bf r}',\omega)={\bar{\bar{\bf G}}}_{0}({\bf r},{\bf r}',\omega)\delta_{fs}+{\bar{\bar{\bf G}}}_{s}^{(fs)}({\bf r},{\bf r}',\omega)$, where ${\bar{\bar{\bf G}}}_{0}$ denotes the contribution of the direct waves from the emitter in an unbounded vacuum, the scattering Green tensor ${\bar{\bar{\bf G}}}_{s}^{(fs)}$ describes the multiple reflection and transmission processes due to the interaction of the emitter with the nanosphere, $f$ and $s$ refer to the regions where the field point and source point are located, and $\delta_{fs}$ is the usual Kronecker delta. In the current study, the atom(emitter) is located out of the nanosphere on the $z$-axis. This, together with the analysis of Eqs.~(\ref{alpha})-(\ref{g_ij}), leads us to the fact that the field (source) point is placed outside (both outside and inside) of the nanosphere. Since, the dipole moment of the atom is along the radial direction, \textrm{i.e.,} $\hat{\bf r}$, only the radial component of the direct term of the Green's function is needed to compute the PLDOS, which is given by:
\begin{eqnarray}\label{direct term}
{\bar{\bar{\bf G}}}_{0,rr}({\bf r}_a,{\bf r}_a,\omega)&=&\frac{i k_0}{4\pi}\sum_{n}n (n + 1)(2 n + 1) \nonumber\\
&&\times \frac{h^{(1)}_n(k_0r_a) j_n(k_0r_a)}{(k_0 r_a)^2}.
\end{eqnarray}
While, for the scattered part, the following components are required in the PLDOS and $\rho_m$ calculations:
\begin{subequations}\label{scattered part}
\begin{eqnarray}
{\bar{\bar{\bf G}}}_{s,rr}^{(00)}({\bf r}_a,{\bf r}_a,\omega)&=&\frac{i k_0}{4\pi}\sum_{n} n (n + 1)(2 n + 1)B_N^{00}(\omega) \,\,\,\,\,\,\,\,\,\,\,\,\,\, \label{scattered part1}\\
&& \times  \Big(\frac{h^{(1)}_n(k_0r_a)}{k_0r_a}\Big)^2 , \nonumber\\
{\bar{\bar{\bf G}}}_{s,rr}^{(01)}({\bf r}_a,{\bf r}',\omega)&=&\frac{i k_1}{4\pi}\sum_{n} n (n + 1)(2 n + 1)A_N^{01}(\omega) \label{scattered part2}\\
&& \times \Big(\frac{h^{(1)}_n(k_0r_a) j_n(k_1r')}{k_0r_a \,k_1r' }\Big) P_n(\cos \theta') ,\nonumber\\
{\bar{\bar{\bf G}}}_{s,r\theta}^{(01)}({\bf r}_a,{\bf r}',\omega)&=&\frac{i k_1}{4\pi}\sum_{n} (2 n + 1)A_N^{01}(\omega) \label{scattered part3}\\
&& \times \Big(\frac{h^{(1)}_n(k_0r_a) \partial j_n(k_1r')}{k_0r_a \,k_1r'  }\Big) \frac {d P_n(\cos \theta')}{d\theta'} ,\nonumber\\
\bar{\bar{G}}_{s,r\varphi}^{(01)}({\bf r}_a,{\bf r}',\omega)&=&0,\label{scattered part4}
\end{eqnarray}
\end{subequations}
where $k_0=\omega/c$, $k_1=\sqrt{\epsilon}\,\omega/c$, and the prime in the last three equations represents the coordinates $(r',\theta',\varphi')$ of the source inside the sphere. Here, $j_n(x)$ is the spherical Bessel function of the first kind, $h^{(1)}_n(x)$ is the first-type spherical Hankel function, and $P_n^m(x)$ is the associated Legendre function.

The scattering coefficients $B_{N}^{00}(\omega)$, and $A^{01}_{N}(\omega)$ in Eqs.~(\ref{scattered part}) are related to the reflection and transition coefficients $-R^{V}_{F0}(\omega)$, and $ T^{V}_{F0}(\omega)$, respectively, with
\begin{subequations}
\begin{eqnarray}\label{refl}
R^V_{F1}&=& \frac{k_1 j_n(k_1 a) \partial j_n( k_0 a) -  k_0 j_n(k_0 a)\partial j_n( k_1 a)}{k_1 j_n(k_1 a)\partial h^{(1)}_n(k_0 a)- k_0 \partial j_n(k_1 a) h^{(1)}_n( k_0 a)},\,\,\,\,\,\,\,\,\,\,\,\,\,\,\\
T^V_{F1}&=& \frac{k_1( s_n(k_1 a) \partial h^{(1)}_n( k_1 a) - \partial j_n(k_1 a)h^{(1)}_n( k_1 a))}{k_1 j_n(k_1 a)\partial h^{(1)}_n(k_0 a)- k_0 \partial j_n(k_1 a) h^{(1)}_n( k_0 a)},\,\,\,\,\,\,\,\,\,\,\,\,\,\,\\\nonumber
\end{eqnarray}
\end{subequations}
where we introduced the abbreviations $\partial j_n(x)= \frac{1}{x}\frac{d(x j_n(x) )}{ dx}$ and $\partial h^{(1)}_n(x)= \frac{1}{x}\frac{d(x h^{(1)}_n(x) )}{dx}$.

Inserting Eqs.~(\ref{direct term}) and~(\ref{scattered part1}) into Eq.~(\ref{alpha}), after some manipulations, the PLDOS is rewritten as:
\begin{eqnarray}
\rho(\hat{\bf r},{\bf r}_a,\omega)&=&\rho_0 \bigg (1+\frac{3}{2}{\rm Re}\sum_{n} n (n + 1)(2 n + 1)B_N^{00}(\omega)\nonumber \\
&&\hspace{-1cm} \times \Big(\frac{h^{(1)}_n(k_0r_a)}{k_0r_a}\Big)^2  \bigg).
\end{eqnarray}
Likewise, substituting Eqs.~(\ref{scattered part2})-(\ref{scattered part4}) into Eq.~(\ref{g_ij}) after lengthy but straightforward calculations, the medium PLDOS $\rho_m$ is simplified as
\begin{equation}
\rho_m(\hat{\bf r},{\bf r}_a,\omega)=\frac{24 \pi \epsilon_0}{\omega}\mathrm {Im}\epsilon(\omega)\big({\cal C}_{1}( {\bf r}_a,\omega) +{\cal C}_{2}({\bf r}_A,\omega)\big),
\end{equation}
where
\begin{subequations}\label{C1C2}
\begin{eqnarray}
{\cal C}_{1}({\bf r}_a,\omega)&=&\sum_{n} n^2(n+1)^2(2n+1)\vert A^{01}_N(\omega)\vert^2 \\
&&\times \Big|\frac{h^{(1)}_n(k_0r_a)}{k_0r_a}\Big|^2 \int_0^a dr' r'^2 \Big|\frac{j_n(k_1r')}{k_1 r'}\Big|^2,\nonumber\\
{\cal C}_{2}({\bf r}_a,\omega)&=&\sum_{n} n(n+1)(2n+1)\vert A^{01}_N(\omega)\vert ^2 \\
&&\times \Big|\frac{h^{(1)}_n(k_0r_a)}{k_0r_a}\Big|^2 \int_0^a dr' r'^2 \vert \partial j_n(k_1r')\vert^2.\nonumber
\end{eqnarray}
\end{subequations}
Here, the following integral identities are used to get Eqs.~(\ref{C1C2}):
\begin{subequations}
\begin{eqnarray}
\int_0^\pi \frac {d P_n(\cos \theta')}{d\theta'} \frac {d P_{n'}(\cos \theta')}{d\theta'} \sin \theta'd \theta'&=&\frac{2n(n+1)}{2n+1}\delta_{nn'},\nonumber\\
\int_0^\pi P_n(\cos \theta') P_{n'}(\cos \theta') \sin \theta'd \theta'&=&\frac{2}{2n+1}\delta_{nn'}.\nonumber
\end{eqnarray}
\end{subequations}


\begin{thebibliography}{10}
%
\bibitem{Pancharatnam1975} S. Pancharatnam, in Collected Works of S. Pancharatnam
(Oxford University Press, Oxford, 1975).
%
\bibitem{Berry} M. V. Berry, Proc. R. Soc. Lond. A \textbf{392}, 45 (1984).
%
\bibitem{Berry1990} M. V. Berry, Phys. Today \textbf{43}(12), 34 (1990).
%
\bibitem{Aharonov1959} Y. Aharonov and D. Bohm, Phys. Rev. \textbf{115}, 485 (1959).
%
\bibitem{Aharonov} Y. Aharonov and J. Anandan, Phys. Rev. Lett. \textbf{58}, 1593 (1987).
%
\bibitem{Anandan1} J. Anandan and Y. Aharonov, Phys. Rev. D \textbf{38}, 1863 (1988).
%
\bibitem{Anandan2} J. Anandan, Phys. Lett. A \textbf{133}, 171 (1988).
%
\bibitem{Samuel} J. Samuel and R. Bhandari, Phys. Rev. Lett. \textbf{60}, 2339 (1988).
%
\bibitem{Pati} A. K. Pati, Phys. Rev. A \textbf{52}, 2576 (1995).
%
%
\bibitem{KleiBler2018} F. Klei{\ss}ler, A. Lazariev, and S. Arroyo-Camejo, npj Quantum Inf \textbf{4}, 49 (2018).
%
\bibitem{Huang2019} Y.-Y. Huang, Y.-K. Wu, F. Wang, P.-Y. Hou, W.-B. Wang, W.-G. Zhang, W.-Q. Lian, Y.-Q. Liu, H.-Y. Wang, H.-Y. Zhang, L. He, X.-Y. Chang, Y. Xu, and L.-M. Duan, Phys. Rev. Lett. \textbf{122}, 010503 (2019).
%
\bibitem{Cho} Y.-W. Cho, Y. Kim, Y.-H. Choi, Y.-S. Kim, S.-W. Han, S.-Y. Lee, S. Moon, and Y.-H. Kim, Nat. Phys. \textbf{15}, 665 (2019).
%
\bibitem{Johnsson} M. T. Johnsson , N. R. Mukty , D. Burgarth , T. Volz, and G. K. Brennen, Phys Rev. Lett. \textbf{125}, 190403 (2020).
%
\bibitem{Chen1} T. Chen and Z.-Y. Xue, Phys. Rev. Applied \textbf{14}, 064009 (2020).
%
\bibitem{Zhang} J. Zhang, T. H. Kyaw, S. Filipp, L.-C. Kwek, E. Sjoqvist, D. Tong, Phys. Rep. \textbf{1027}, 1 (2023).
%
\bibitem{Colmenar} R. K. L. Colmenar , U. Gungordu,  and J. P. Kestner, PRX Quantum \textbf{3}, 030310 (2022).
%
\bibitem{Whitney1} R. S. Whitney and Y. Gefen, Phys. Rev. Lett. \textbf{90}, 190402 (2003).
%
\bibitem{Whitney2} R. S. Whitney, Y. Makhlin, A. Shnirman, and Y. Gefen, Phys. Rev. Lett. \textbf{94}, 070407 (2005).
%
\bibitem{Carollo1} A. Carollo, I. Fuentes-Guridi, M. F. Santos, and V. Vedral, Phys. Rev. Lett. \textbf{90}, 160402 (2003).
%
\bibitem{Carollo2} A. Carollo. Mod. Phys. Lett. A \textbf{20}, 1635 (2005).
%
\bibitem{Tong} D. M. Tong, E. Sjoqvist, L. C. Kwek, and C. H. Oh, Phys. Rev. Lett. \textbf{93}, 080405 (2004).
%
\bibitem{Marzlin} K. P. Marzlin, S. Ghose, and B. C. Sanders, Phys. Rev. Lett. \textbf{93}, 260402 (2004).
%
\bibitem{Lombardo} F. C. Lombardo and P. I. Villar, Phys. Rev. A \textbf{74}, 042311 (2006).
%
\bibitem{Villar1} P. I. Villar and F. C. Lombardo, Phys. Rev. A \textbf{83}, 052121 (2011).
%
\bibitem{Lombardo2} F. C. Lombardo and P. I. Villar, Int. J. of Quantum Information \textbf{6}, 707713 (2008).
%
\bibitem{Lombardo3} F. C. Lombardo and P. I. Villar, Phys. Rev. A \textbf{81}, 022115 (2010).
%
\bibitem{Lombardo4} F. C. Lombardo and P. I. Villar, Phys. Rev. A \textbf{87}, 032338 (2013).
%
\bibitem{Villar2} P. I. Villar and A. Soba, Phys. Rev. A \textbf{101}, 052112 (2020).
%
\bibitem{hu} J. Hu and H. Yu, 	Phys. Rev. A \textbf{85}, 032105 (2012).
%
\bibitem{Chen} J. J. Chen, J. H. An, Q. J. Tong, H. G. Luo, and C. H. Oh, Phys. Rev. A \textbf{81}, 022120 (2010).
%
\bibitem{Dechiara} G. De Chiara and G. M. Palma, Phys. Rev. Lett. \textbf{91}, 090404 (2003).
%
\bibitem{Banerjee} S. Banerjee and R. Srikanth, Eur. Phys. J. D \textbf{46}, 335 (2008).
%
\bibitem{Yi} X. X. Yi, L. C. Wang, and W. Wang, Phys. Rev. A \textbf{71}, 044101 (2005).
%
\bibitem{Wang} Z. S. Wang, C. Wu, X.-L. Feng, L. C. Kwek, C. H. Lai, and C. H. Oh, Phys. Rev. A \textbf{75}, 024102 (2007).
%
\bibitem{Cai2019} X. Cai, R. Meng, Y. Zhang, and L. Wang, Europhys. Lett. \textbf{125}, 30007 (2019).
%
\bibitem{Villar2022} L. Viotti, F. C. Lombardo, and P. I. Villar, Phys. Rev. A \textbf{105}, 022218 (2022).
%
\bibitem{Fuentes-Guridi1} I. Fuentes-Guridi, A. Carollo, S. Bose, and V. Vedral, Phys. Rev. Lett. \textbf{89}, 220404 (2002).
%
\bibitem{Fuentes-Guridi2} I. Fuentes-Guridi, S. Bose, and V. Vedral, Phys. Rev. Lett. \textbf{85}, 5018 (2000).
%
\bibitem{Martin-Martinez1} E. Martin-Martinez, I. Fuentes, and R. B. Mann,  Phys. Rev. Lett. \textbf{107}, 131301 (2011).
%
\bibitem{Farias} M. B. Farias, F. C. Lombardo, A Soba, P. I. Villar and R. S. Decca, npj Quantum Inf \textbf{6}, 25 (2020).
%
\bibitem{Martin-Martinez2} E. Martin-Martinez, A. Dragan, R. B. Mann, and I. Fuentes, New J.  Phy.  \textbf{15}, 053036 (2013).
%
\bibitem{Bellomo1} B. Bellomo, R. Messina, and M. Antezza, Europhys. Lett. \textbf{100}, 20006 (2012).
%
\bibitem{Bellomo2} B. Bellomo, R. Messina, D. Felbacq, and M. Antezza, Phys. Rev. A \textbf{87},012101 (2013).
%
\bibitem{Cai2016} X. Cai and Y. Zheng, Phys. Rev. A \textbf{94}, 042110 (2016).
%
\bibitem{Milonni} P. W. Milonni, The Quantum vacuum (Academic, San Diego, 1994).
%
\bibitem{Huttner} B. Huttner and S. M. Barnett, Phys. Rev. A \textbf{46}, 4306 (1992).
%
\bibitem{Jeffers} J. Jeffers, S.M. Barnett, R. Loudon, R. Matloob, and M. Artoni, Opt. Commun. \textbf{131}, 66 (1996).
%
\bibitem{Suttorp} L. G. Suttorp and M. Wubs, Phys. Rev. A, \textbf{70}, 013816 (2004).
%
\bibitem{Amooshahi} M. Amooshahi, J. Math. Phys. \textbf{50}, 062301 (2009).
%
\bibitem{Kheirandish1} F. Kheirandish, E. Amooghorban, and M. Soltani, Phys. Rev. A \textbf{83}, 032507 (2011).
%
\bibitem{Philbin} T. G. Philbin, New J. Phys. \textbf{12}, 123008 (2010).
%
\bibitem{Kheirandish2} F. Kheirandish, and E. Amooghorban, Phys. Rev. A \textbf{82}, 042901 (2010).
%
\bibitem{Amooghorban} E. Amooghorban, M. Wubs, N. A. Mortensen, and F. Kheirandish, Phys. Rev. A \textbf{84}, 013806 (2011).
%
\bibitem{Gruner} T. Gruner and D.-G. Welsch, Phys. Rev. A \textbf{53}, 1818 (1996).
%
\bibitem{Dung} H. T. Dung, L. Kn\"{o}ll, and D.-G. Welsch, Phys. Rev. A \textbf{57}, 3931 (1998).
%
\bibitem{Scheel} S. Scheel, L. Kn\"{o}ll, and D.-G. Welsch,  Phys. Rev. A  \textbf{58}, 700 (1998).
%
\bibitem{Matloob1} R. Matloob, R. Loudon, S. M. Barnett, and J. Jeffers, Phys. Rev. A \textbf{52}, 4823 (1995).
%
\bibitem{Knoll}  L. Kn\"{o}ll, S. Scheel, and D.-G. Welsch, Coherence and Statistics of Photons and Atoms (Wiley, New York, 2001).
%
\bibitem{Matloob2} R. Matloob, Phys. Rev. A  \textbf{70}, 022108 (2004).
%
\bibitem{Behbahani} M. Morshed Behbahani, E. Amooghorban, and A. Mahdifar, Phys. Rev. A \textbf{94}, 013854 (2016).
%
\bibitem{Amooghorban17} E. Amooghorban, and E. Aleebrahim, Phys. Rev. A \textbf{96}, 012339 (2017).
%
\bibitem{Breuer} H.-P. Breuer and F. Petruccione, The Theory of Open Quantum Systems (Oxford University Press, New York, 2002).
%
\bibitem{Hu} J. Hu, W. Zhou, and H. Yu, Phys. Rev. D \textbf{88}, 085035 (2013).
%
\bibitem{Wu} P. Wu, and H. Yu, Phys. Rev. A \textbf{92}, 062503 (2015).
%
\bibitem{Novotny2006}   L. Novotny and B. Hecht, Principles of Nano-optics (Cambridge
University Press, Cambridge, 2006).
%
\bibitem{Leek2007} P. J. Leek, J. M. Fink, A. Blais, R. Bianchetti, M. G\"{o}ppl,
J. M. Gambetta, D. I. Schuster, L. Frunzio, R. J. Schoelkopf, and
A. Wallraff, Science \textbf{318} , 1889 (2007).
%
\bibitem{Palik} Handbook of Optical Constants of Solids, edited by E. Palik. (Academic Press, New York, 1998).
%
\bibitem{Kittle1987} C. Kittle, Quantum Theory of Solids (Wiley, New York, 1987).
%
\bibitem{Zibik} E. A. Zibik, T. Grange, B. A. Carpenter, N. E. Porter, R. Ferreira, G. Bastard, D. Stehr, S. Winnerl, M. Helm, H. Y. Liu, M. S. Skolnick and L. R. Wilson, Nat. Mater. \textbf{8}, 803 (2009).
%
\bibitem{Li1994} L. W. Li, P. S. Kooi, M. S. Leong, and T. S. Yeo, IEEE Trans.
Microwave Theory Tech. \textbf{42}, 2302 (1994).
%
%
\end{thebibliography}
 \end{document}